\documentclass[aps,preprint]{revtex4}
\usepackage{graphics,amsmath,amsfonts}

\begin{document}
\title{Homogenization of membrane and pillar photonic crystals}
\author{Didier Felbacq}
\affiliation{GES UMR 5650 - Bat. 21 CC074 Place E. Bataillon 34095 Montpellier Cedex 05, France}

\author{Guy Bouchitt\'e}
\affiliation{ANLA Universit\'e de Toulon BP 132 - 83957 La Garde Cedex, France}

\author{Antoine Moreau}

\affiliation{LASMEA UMR 6602 - Complexe des C\'ezeaux 63177 Aubi\`ere Cedex, France}

\begin{abstract}
We study wave propagation and diffraction in a bidimensional photonic crystal
with finite height, in case where the wavelength is large with respect to the period of
the structure. The device is made of materials with anisotropic 
permittivity and permeability tensors. We derive rigorously the homogenized system, using the
concept of two-scale convergence. The effective permittivity and
permeability tensors turn out to be that of a two-dimensional photonic
crystal with infinite height.
\end{abstract}


\maketitle
\date{\today}

Photonic crystals, i.e. dielectric or metallic artificial periodic
structures, are generally thought of as strongly scattering devices,
authorizing the existence of photonic band gaps. However, their actual
electromagnetic behavior when the wevelength is large with respect to the
period is also interesting, because it can produce strongly anisotropic
behaviors, plasmon frequencies, or even left-handed materials \cite{pendry1,pendry2}. The study of
the properties of these structures in this asymptotic regime comes under
the theory of homogenization \cite{kozlov}. A lot of results are by now very well-known both for
2D and 3D structures. In this paper, we consider a photonic crystal made of
a collection of parallel finite-size fibers embedded in a matrix. 
This covers the case of structures made out of
a layer of bulk materials in which holes are
made periodically (membrane photonic crystal)
 but also the case of structures made out of nanopillars (pillar photonic crystal
\cite{pobor,tada,peyr,monat,francois,benisty}), or more generally, structures composed
of fibers with finite length embedded in a matrix.
Our point is to derive the effective permittivity and permeability tensors of this
structure when the ratio between the period of the structure and the
wavelength of the incident field is very small.
We show, using the two-scale convergence method, that the effective, or homogenized,
permittivity and permeability tensors of these structures are the same as that of infinitely long fibers, 
for which we had already derived rigorous results \cite{moibou}. 
For infinitely long fibers, explicit formulas can be derived in some cases \cite{kozlov,papa,mcphed1,mcphed2}.
Let us note that our results hold for dispersive and lossy materials.

 The set of fibers is contained in a domain $\Omega =\mathcal{O}\times \left[
-L,L\right] $ of $\Bbb{R}^{3}$ (cf. fig.1). The space coordinates are denoted:
 $\mathbf{x}=\left( x_1,x_2,x_3 \right)$ and we also denote 
$\mathbf{x}_\bot=\left( x_1,x_2 \right)$. The period of the lattice is
denoted by $\eta $ (see fig. 2). We denote by $Y$ the basic two-dimensional 
cell of the lattice. The obstacle in $Y$ is denoted by $P$. 
We consider time harmonic fields, the time dependence is chosen to be $\exp \left( -i\omega t\right) $.
For a given monochromatic incident field 
$\left( \mathbf{E}^{i},\mathbf{H}^{i}\right) $, we denote by 
$\left( \mathbf{E}^{\eta },\mathbf{H}^{\eta }\right) $ 
the total electromagnetic field. Our aim is
to pass to the limit $\eta \rightarrow 0$ and determine the limit of the
couple $\left( \mathbf{E}^{\eta },\mathbf{H}^{\eta }\right) $. In our
methodology, we get at the limit a true electromagnetic scattering problem,
for a given wavelength $\lambda $ and a bounded obstacle $\Omega $
characterized by some permittivity and permeability tensors. This situation
is quite different from other homogenization techniques, making use of periodization conditions,
 in which the frequency tends to zero, thus
not leading to a diffraction problem but rather to an electrostatic one \cite{arriaga}.
Such an approach can sometimes give useful explicit formulas but generally
leads to complicated formulations. Moreover, it does not handle the boundary
effects which in some cases may lead to some miscomprehensions \cite{moi}.
The relative permittivity tensor 
$\varepsilon^{\eta }\left( \mathbf{x}\right) $ and relative permeability tensor 
$\mu^{\eta }\left( \mathbf{x}\right) $ are described by : 
\begin{equation}
\left\{ 
\begin{array}{ll}
\varepsilon ^{\eta }\left( \mathbf{x}\right) =\varepsilon _{0} & 
\text{for } \mathbf{x}\in \Bbb{R}^{3}\backslash \Omega \\ 
\varepsilon ^{\eta }\left( \mathbf{x}\right) =\varepsilon ^{0}
\left( \frac{\mathbf{x}_\bot}{\eta },x_3\right) & \text{for }\mathbf{x}_\bot\in \Omega
\end{array}
\right. ,\left\{ 
\begin{array}{ll}
\mu ^{\eta }\left( \mathbf{x}\right) =\mu _{0} & \text{for }\mathbf{x}\in 
\Bbb{R}^{3}\backslash \Omega \\ 
\mu ^{\eta }\left( \mathbf{x}\right) =\mu ^{0}
\left( \frac{\mathbf{x}_\bot}{\eta },x_3\right) & \text{for }\mathbf{x}_\bot\in \Omega
\end{array}
\right.
\end{equation}
where $\mathbf{y}\rightarrow \varepsilon ^{0}\left( \mathbf{y}\right)
=\left( \varepsilon _{ij}^{0}\left( \mathbf{y}\right) \right) $ and 
$\mathbf{y}\rightarrow \mu ^{0}\left( \mathbf{y}\right) =
\left( \mu _{ij}^{0}\left( \mathbf{y}\right) \right) $ 
are $Y$-periodic $3\times 3$ matrix functions.
The domain $\Omega $ is periodically filled with contracted cells 
$\eta Y\times \left[ -L,L\right] $ (see fig. 2).

The total electromagnetic field 
$\left( \mathbf{E}^{\eta },\mathbf{H}^{\eta}\right) $ satisfies 
\begin{equation}
\left\{ 
\begin{array}{l}
\text{curl }\mathbf{E}^{\eta }=i\omega \mu _{0}\mu ^{\eta }\mathbf{H}^{\eta }
\\ 
\text{curl }\mathbf{H}^{\eta }=-i\omega \varepsilon _{0}\varepsilon ^{\eta }%
\mathbf{E}^{\eta }
\end{array}
\right.
\end{equation}
\newline
and 
$\left( \mathbf{E}^{\eta }-\mathbf{E}^{i},\mathbf{H}^{\eta }-\mathbf{H}^{i}\right) $ 
satisfies Silver-M\"{u}ller radiation conditions.

In order to describe this problem, we will rely on a two-scale expansion of
the fields. That way, the physical problem is described by two variables:
a macroscopic one $\mathbf{x}$ and a microscopic one $\mathbf{y}$
representing the rapid variations of the material at the scale
of one basic cell, measured by $\eta$, that is, at the scale of $\eta Y$. 
By noticing that there are no rapid variations in the vertical direction
$x_3$, the microscopic variable is set to be: $\mathbf{y}=\mathbf{x}_{\bot}/\eta$. 
Differential operators with
respect to variable $\mathbf{y}$ are denoted with a subscript $y$.
The fields are periodic with respect to that
microscopic variable (this corresponds to the neighborhood of the $\Gamma $
point in the first Brillouin zone). The limit problem obtained by letting 
$\eta $ tend to $0$, will then depend on the macroscopic, physical, variable 
$\mathbf{x}$ but also on the microscopic, hidden, variable $\mathbf{y}$. The
total limit fields will read $\mathbf{E}^{0}\left( x,y\right) $ and $\mathbf{%
H}^{0}\left( x,y\right) $ and the observable, physical, fields will be given
by the mean value over the hidden variable $\mathbf{y}$: 
$\mathbf{E}\left(x\right) =
\left| Y\right| ^{-1}
\int_{Y}\mathbf{E}^{0}\left( x,y\right) dy$
and 
$\mathbf{H}\left( x\right) 
=\left| Y\right| ^{-1}\int_{Y}\mathbf{H}^{0}\left( x,y\right) dy$,
where $\left| Y\right|$ is the measure area of $Y$.
 In order to lighten the notations, we denote by
brackets the averaging over $Y$, hence $\mathbf{H}\left( x\right)
=\left\langle \mathbf{H}^{0}\right\rangle $ and $\mathbf{E}\left( x\right)
=\left\langle \mathbf{E}^{0}\right\rangle $.

The main mathematical tool that we use is a mathematically clean version of
the multiscale expansion, widely used in various areas of physics. More
precisely, for a vector field $\mathbf{F}^{\eta }$ in $\left( L^{2}\left(
\Omega \right) \right) ^{3}$, we say, by definition, that $\mathbf{F}^{\eta
} $ two-scale converges towards $\mathbf{F}^{0}$ if
for every sufficiently regular function $\mathbf{\phi }
\left( \mathbf{x},
\mathbf{y}\right) $,
 $Y$-periodic with respect to $\mathbf{y}$, we have: 
\begin{equation}
\int_{\Omega }
\mathbf{E}^{\eta }\left( \mathbf{x}\right) .
\mathbf{\phi }\left(\mathbf{x},\mathbf{x}_\bot/\varepsilon \right) 
dx
\rightarrow 
\iint_{\Omega \times Y}
\mathbf{E}^{0}\left( \mathbf{x},\mathbf{y}\right) .
\mathbf{\phi }\left( \mathbf{x},\mathbf{y}\right) 
dxdy\text{,}
\label{twoscale}
\end{equation}
as $\eta $ tends to $0$.

The limit field $\mathbf{F}^{0}$ is square
integrable with respect to both variables $\mathbf{x}$ and $\mathbf{y}$ and
is $Y$-periodic in the $\mathbf{y}$ variable (it belongs to 
$L^{2}\left( \Omega ;\left( L_{\#}^{2}\left(Y\right) \right) ^{3}\right) $). 
A complete analysis of this
new mathematical tool can be found in \cite{allaire}.

We make the physically reasonable assumption that the electromagnetic energy
remains bounded when $\eta$ tends to $0$, which is equivalent to assume that
 $\left( \mathbf{E}^{\eta},\mathbf{H}^{\eta }\right) $
 are both locally square integrable. Then it is
known \cite{allaire} that $\left( \mathbf{E}^{\eta },\mathbf{H}^{\eta }\right) $
two-scale converges towards limit fields 
$\left( \mathbf{E}^{0},\mathbf{H}^{0}\right) $. 
This physical assumption could be justified mathematically, however
it seems quite obvious, from the point of view of physics, that the limit fields
exist. The point is then to give the system of equations that is
satisfied by these fields and to derive the effective permittivity and permeability
tensors.

First of all, we have to determine the set of equations that are
microscopically satisfied, that is, satisfied with respect 
to the hidden variable $\mathbf{y}$, for that will give the
 constitutive relations of the homogenized medium.
Multiplying Maxwell-Faraday equation by a regular test function 
$\mathbf{\phi }\left( \mathbf{x},\frac{\mathbf{x}_\bot}{\eta }\right) $, 
and integrating over $\Omega $, we obtain: 
\begin{equation}
\int_{\Omega }\mathbf{E}^{\eta }\left( \mathbf{x}\right) .
 \left[ \text{curl}_{x}\left( \mathbf{\phi }\right) 
+\frac{1}{\eta }\text{curl}_{y}\left(\mathbf{\phi }\right) \right] 
d\mathbf{x}=
i\omega \mu _{0}\int_{\Omega }\mu^{\eta }
\left( \mathbf{x}\right) \mathbf{H}^{\eta }\left( \mathbf{x}\right) 
\mathbf{\phi }\left( \mathbf{x},\mathbf{x}_\bot/\eta\right) d\mathbf{x}\text{.}
\end{equation}
Multiplying by $\eta $ and letting $\eta $ tend to $0$, we get using (\ref
{twoscale}): 
\begin{equation}
\iint_{\Omega \times Y}
\mathbf{E}^{0}\left( \mathbf{x},\mathbf{y}\right) .
\text{curl}_{y}\left( \mathbf{\phi }\right) d\mathbf{x}d\mathbf{y}=0.
\end{equation}

 This is equivalent to:
\begin{equation}
\iint_{\Omega \times Y}
\text{curl}_{y}\mathbf{E}^{0}\left( \mathbf{x},\mathbf{y}\right) .
 \mathbf{\phi }\left( \mathbf{x},\mathbf{y}\right) d\mathbf{x}d\mathbf{y}=0
\end{equation}
which is nothing else but the variational form for: curl$_{y}\mathbf{E}^{0}=0$. In a very similar way, but using now
Maxwell-Ampere equation, we get the equation: curl$_{y}\mathbf{H}^{0}=0$. 
On the other hand, since $\varepsilon^{\eta} \mathbf{E}^{\eta}$ is divergence free, we have,
for every test function $\phi(\mathbf{x},\mathbf{y})$, 
$\int_{\Omega }
\varepsilon ^{\eta }\left(\mathbf{x}\right) 
\mathbf{E}^{\eta }\left( \mathbf{x}\right) . 
\left[
\nabla _{x}\phi +\frac{1}{\eta }\nabla _{y}\phi 
\right] 
 d\mathbf{x}=0$.
Multiplying by $\eta $ and letting $\eta $ tend to $0$, we get:
\begin{equation}
\iint_{\Omega\times Y}
\varepsilon ^{\eta }\left( \mathbf{y}\right) 
\mathbf{E}^{0}\left(\mathbf{x},\mathbf{y}\right) .\nabla _{y}\phi 
d\mathbf{x}d\mathbf{y}=0,
\end{equation}
this can be written as (notice that the div$_y$ operator acts only on the 
transverse components):
\begin{equation}
\text{div}_{y}\left( \varepsilon ^{0}\mathbf{E}^{0}\right) =0.
\label{dive}
\end{equation}
Similarly, since the magnetic field is divergence free, we derive:
\begin{equation}
\text{div}_{y}\left( \mu^{0}\mathbf{H}^{0}\right) =0.
\label{divh}
\end{equation}
Summing up, we have the following microscopic equations, holding 
inside the basic cell $Y$: 
\begin{equation}
\left\{ 
\begin{array}{l}
\text{div}_{y}\left( \mu ^{0}\mathbf{H}^{0}\right) =0 \\ 
\text{curl}_{y}\mathbf{H}^{0}=0
\end{array}
\right. ,\left\{ 
\begin{array}{l}
\text{div}_{y}\left( \varepsilon ^{0}\mathbf{E}^{0}\right) =0 \\ 
\text{curl}_{y}\mathbf{E}^{0}=0
\end{array}
\right.
\label{micro}
\end{equation}
The systems in (\ref{micro}) are respectively of electrostatic and magnetostatic types. 
This means that, with respect to the hidden variable $\mathbf{y}$, the electric field
and magnetic field satisfy the static Maxwell system. This property is true
only at that scale and not at the macroscopic scale. However, it is these
static equations that determine the effective permittivity and permeability.
Indeed let us analyze this system starting with the electric field. From
the curl relation, we get $\nabla_{y}E_{3}^{0}=0$, and so 
$E_{3}^{0}(\mathbf{x},\mathbf{y}) \equiv E_{3}\left( \mathbf{x}\right) $. Besides, the basic
cell having the geometry of a torus, we get the existence of a regular
periodic function $w_{E}\left( \mathbf{y}\right) $ such that:
\begin{equation} 
\mathbf{E}_{\bot }^{0}=\mathbf{E}_{\bot }+\nabla _{y}w_{E}.
\label{homol}
\end{equation}
The function $w_E$ is the electrostatic potential
associated with the microscopic electrostatic problem. Inserting (\ref{homol}) in
 equation (\ref{dive}) and projecting on the two horizontal axis, we obtain:
\begin{equation} 
\text{div}_{y}\left[ 
\varepsilon ^{0}\left( \mathbf{e}_{i}+\nabla _{y}w_{E,i}\right) 
\right] =0,\text{ }i\in \left\{ 1,2\right\}  
\label{annexee}
\end{equation}
By linearity, denoting $\mathbf{E}_\bot=\left( E_1,E_2 \right)$, we derive that
 the potential $w_E$ is given by $w_{E}=E_{1}w_{E,1}+E_{2}w_{E,2}$, where $w_{E,i}$
 are the periodic solutions of (\ref{annexee}). Thus by (\ref{homol}): 
\begin{equation}
\mathbf{E}^{0}\left( \mathbf{x},\mathbf{y}\right) 
=\mathcal{E}
\left(\mathbf{y}\right) \mathbf{E}\left( \mathbf{x}\right)
\label{el}
\end{equation}
where: 
\begin{equation}
\mathcal{E}\left( \mathbf{y}\right) =\left( 
\begin{array}{ccc}
1+\partial _{y_{1}}w_{E,1} & \partial _{y_{1}}w_{E,2} & 0 \\ 
\partial _{y_{2}}w_{E,1} & 1+\partial _{y_{2}}w_{E,2} & 0 \\ 
0 & 0 & 1
\end{array}
\right)
\label{tensore}
\end{equation}
The magnetic field $\mathbf{H}^0$ can be handled in the same way since it
satisfies exactly the same kind of equations as $\mathbf{H}^0$ (see (\ref{micro})).
In particular, we may represent its tranversal component in the form: 
$\mathbf{H}^0_{\bot}=\mathbf{H}_{\bot}+\nabla_{\bot}w_H$, where $w_H$ is 
a periodic magnetic potential (the possibility of this representation is due
to the curl-free condition which means that no microscopic current is present).
Analogously as in (\ref{el},\ref{tensore}), we find:
\begin{equation}
\mathbf{H}^{0}\left( \mathbf{x},\mathbf{y}\right) =
\mathcal{M}\left( \mathbf{y}\right) 
\mathbf{H}\left( \mathbf{x}\right)  \label{mag}
\end{equation}
with
\begin{equation}
\mathcal{M}\left( \mathbf{y}\right) =\left( 
\begin{array}{ccc}
1+\partial _{y_{1}}w_{H,1} & \partial _{y_{1}}w_{H,2} & 0 \\ 
\partial _{y_{2}}w_{H,1} & 1+\partial _{y_{2}}w_{H,2} & 0 \\ 
0 & 0 & 1
\end{array}
\right)
\end{equation}
where: 
\begin{equation}
\text{div}_{y}\left[ \mu ^{0}\left( \mathbf{e}_{i}+\nabla _{y}w_{H,i}\right) 
\right] =0,\text{ }i\in \left\{ 1,2\right\}  \label{annexeh}
\end{equation}
The above results show that, at the microscopic scale, the limit fields
$(\mathbf{E^0},\mathbf{H^0})$ are
completely determined by the physical fields $(\mathbf{E},\mathbf{H})$.
Now that the microscopic behavior is precised, we are able to determine the 
macroscopic system satisfied by $(\mathbf{E},\mathbf{H})$. To that aim, let us
choose a regular test function $\mathbf{\phi }\left( x\right) $ independent
of variable $\mathbf{y}$. From Maxwell equations we get 
\begin{equation}
\left\{ 
\begin{array}{c}
\int_{\Omega }\mathbf{H}^{\eta }\left( \mathbf{x}\right).
\text{curl}\left( \mathbf{\phi }\right) d\mathbf{x}
=-i\omega \varepsilon
_{0}\int_{\Omega }\varepsilon ^{\eta }\left( \mathbf{x}\right) 
\mathbf{E}^{\eta }\left( \mathbf{x}\right) \mathbf{\phi }
\left( \mathbf{x}\right) d\mathbf{x} \\ 
\int_{\Omega }\mathbf{E}^{\eta }\left( \mathbf{x}\right) .
 \text{curl}\left( \mathbf{\phi }\right) d\mathbf{x}
=i\omega \mu _{0}\int_{\Omega }\mu
^{\eta }\left( \mathbf{x}\right) \mathbf{H}^{\eta }\left( \mathbf{x}\right) 
\mathbf{\phi }\left( \mathbf{x}\right) d\mathbf{x}
\end{array}
\right.
\end{equation}
\newline
passing to the limit $\eta \rightarrow 0$, we get: 
\begin{equation}
\left\{ 
\begin{array}{c}
\iint_{\Omega \times Y}
\mathbf{H}^{0}\left( \mathbf{x},\mathbf{y}\right)
. \text{curl}\left( \mathbf{\phi }\right) d\mathbf{x}d\mathbf{y}
=-i\omega \varepsilon _{0}
\iint_{\Omega \times Y}\varepsilon ^{0}
\left( \mathbf{y}\right) \mathbf{E}^{0}\left( \mathbf{x},\mathbf{y}\right) 
\mathbf{\phi }\left( \mathbf{x}\right) d\mathbf{x}d\mathbf{y} \\ 
\iint_{\Omega \times Y}
\mathbf{E}^{0}\left( \mathbf{x},\mathbf{y}\right)
. \text{curl}\left( \mathbf{\phi }\right) d\mathbf{x}d\mathbf{y}=
i\omega \mu _{0}
\iint_{\Omega \times Y}\mu ^{0}\left( \mathbf{y}\right) 
\mathbf{H}^{0}\left( \mathbf{x},\mathbf{y}\right) \mathbf{\phi }
\left( \mathbf{x}\right) d\mathbf{x}d\mathbf{y}
\end{array}
\right.
\end{equation}
Recalling that $\langle \mathbf{E}^{0} \rangle= \mathbf{E}$ and that
$\langle \mathbf{H}^{0} \rangle= \mathbf{H}$, we get: 
\begin{equation}
\left\{ 
\begin{array}{l}
\text{curl }\mathbf{E}=i\omega \mu _{0} 
\left \langle \mu ^{0} \mathbf{H}^{0}\right \rangle \\ 
\text{curl }\mathbf{H}=-i\omega \varepsilon _{0}
\left \langle \varepsilon^{0} \mathbf{E}^{0} \right \rangle
\end{array}
\right.  \label{homog1}
\end{equation}
\newline
which, taking into account (\ref{el},\ref{mag}), brings to the limit system: 
\begin{equation}
\left\{ 
\begin{array}{l}
\text{curl }\mathbf{E}=i\omega \mu _{0}
\left\langle \mu ^{0} \mathcal{M}\right\rangle \mathbf{H} \\ 
\text{curl }\mathbf{H}=-i\omega \varepsilon _{0}
\left\langle \varepsilon ^{0} \mathcal{E}\right\rangle \mathbf{E}
\end{array}
\right.
\end{equation}

The homogenized permeability and permittivity tensors
are thus respectively $\left\langle \mu ^{0} \mathcal{M}\right\rangle$
and $\left\langle \varepsilon ^{0} \mathcal{E}\right\rangle$.
It appears here that the homogenization process is purely local, and that
the finiteness of the fibers does not play any role in the homogeneous properties
of the medium: the effective tensors coincide with that obtained in the polarized
cases \cite{kozlov,moibou,zolla}. This surprising property
is easily foreseen by our methodology. An approach relying on explicit
calculations, for instance using Bloch-waves theory or Fourier-Bessel
expansions, cannot work here, due the lack of an explicit representation of
the fields in case of finite size fibers. It should also be
noted that the case of materials with losses is handled by our
result. This result can be straightforwardly applied to the study of
membrane photonic crystal in the long wavelength range where phenomena of
birefringence and dichroism are obtained \cite{pekin}. However, we emphasize 
that the locality pointed out, that is, the fact the effective constitutive
relations are local ones, is lost if we change the scale of the
permittivity coefficients in the obstacles. In particular, the results obtained
in the case of infinite conductivities in the polarized case \cite{moibou} cannot
be transposed to the case of fibers with finite length, due to the emergence of
surprising non local effects which are studied in \cite{metal1,metal2}.
We also remark that the
situation that we handle here is different from that studied in \cite
{lalanne} where the small parameter is the depth over wavelength, while the
period over wavelength ratio is not small, contrarily to our situation. In
that case, a dependence on the depth is found. In our homogenization result,
it is clear that the main numerical problem is the solving of the annex
problems (\ref{annexee},\ref{annexeh}) for they give the effective matrices 
$\mathcal{E}$ and $\mathcal{M}$. In certain simple cases, for instance that
of circular isotropic non magnetic rods and a permittivity constant in each
connected region, it is possible to find an explicit expression for the
effective permittivity (it is in fact a very old problem). However, for more
complicated geometries, there is a general numerical procedure based on
fictitious sources, that allows to solve both annex problems at a low
numerical cost \cite{mfs}.

\newpage {\large {Figures captions} }

Figure 1: Schematics of the photonic crystal 

Figure 2: Schematics of the basic cells. 

(a) Tridimensional basic cell with cylindrical obstacle.

(b) Bidimensional basic cell $Y$ with 2D obstacle $P$.

 \begin{figure}[h]
 \begin{center}
    \includegraphics{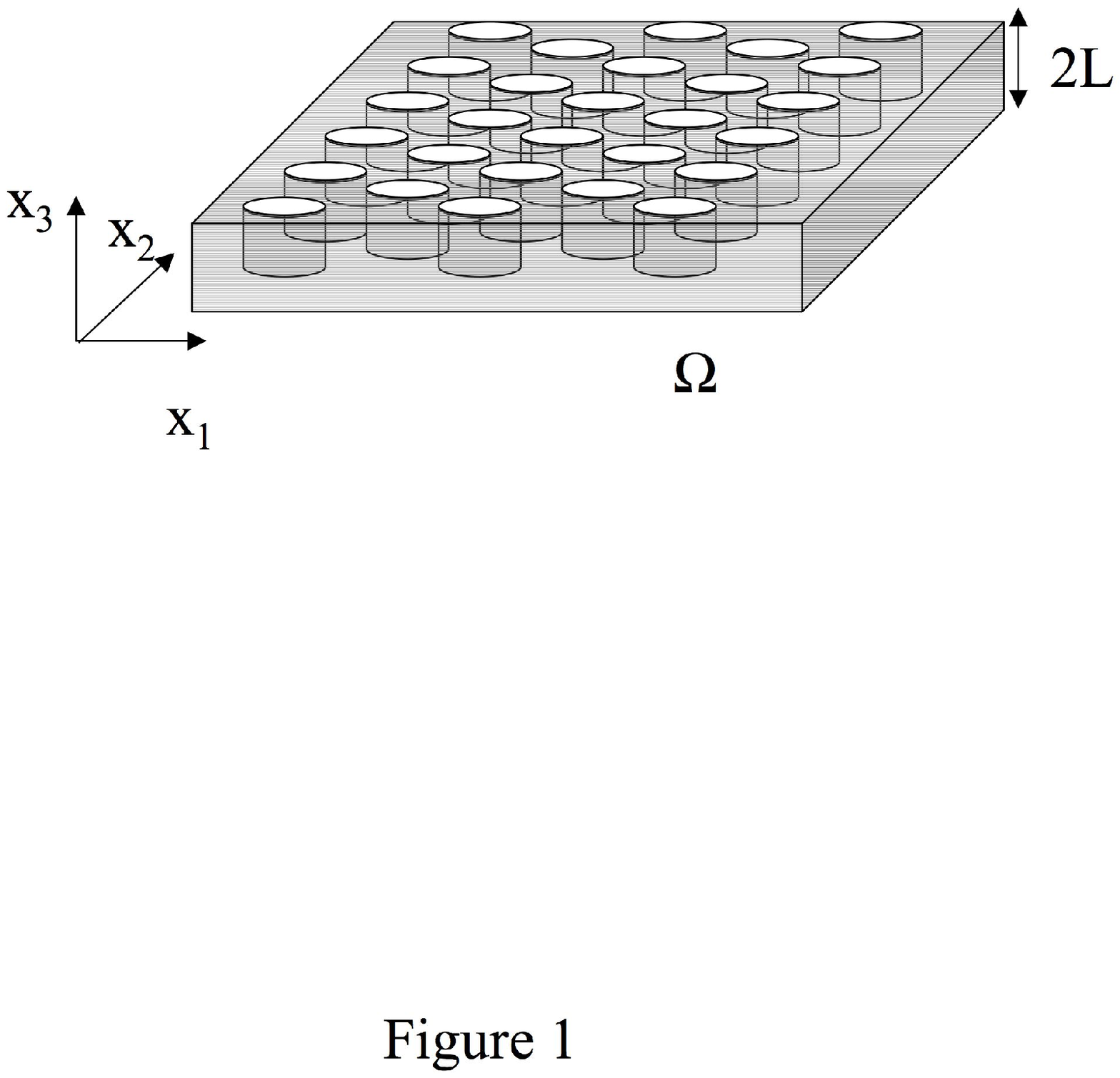}
    \end{center}
     \end{figure}
\begin{figure}[h]
 \begin{center}
    \includegraphics{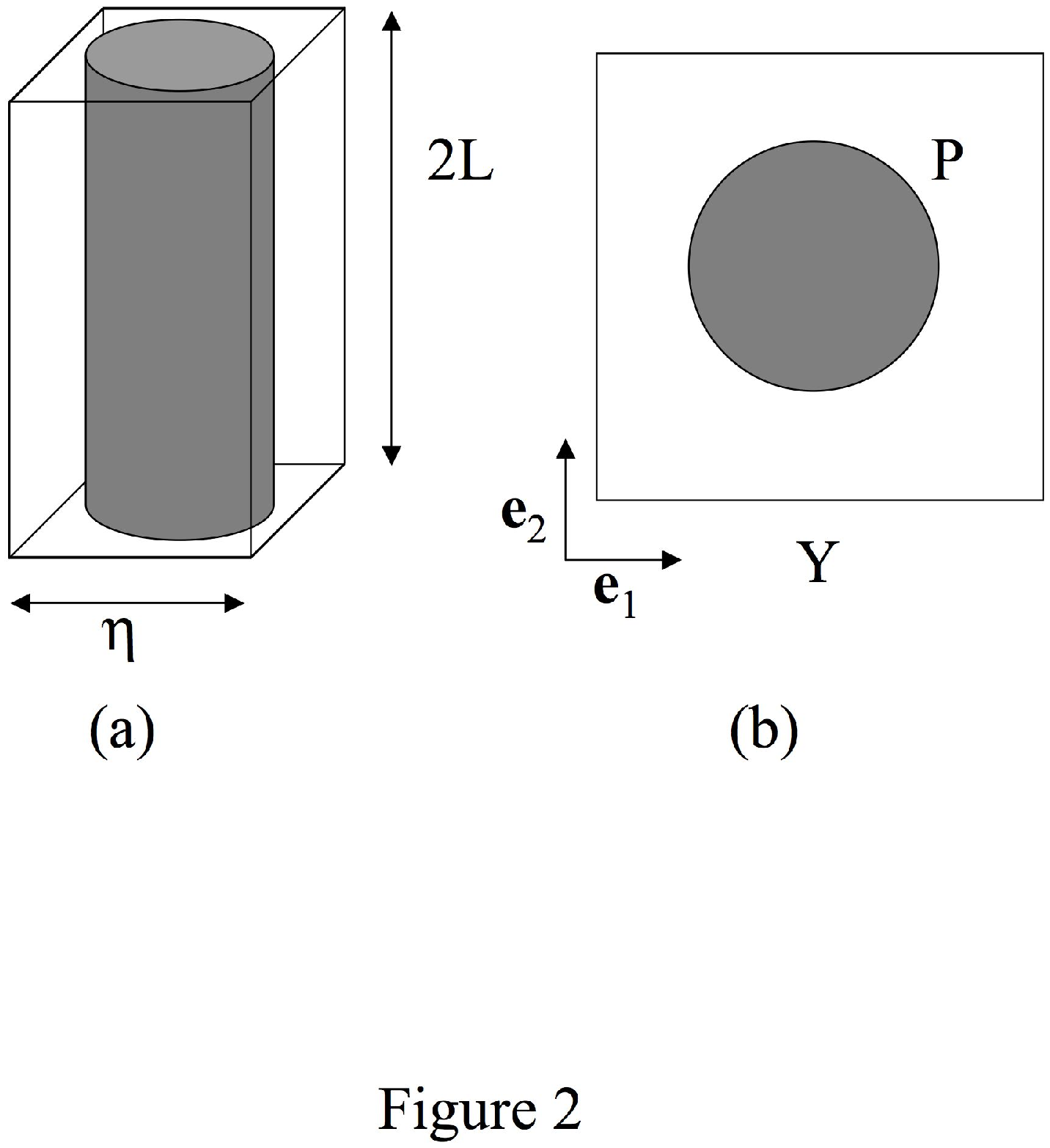}
    \end{center}
     \end{figure}

\end{document}